# Atomic Origins of Magnetic Anisotropy in Ru-substituted Manganite Films

Brajagopal Das,[1,a)] Cinthia Piamonteze,[2] Lior Kornblum[1]

[1]Andrew & Erna Viterbi Department of Electrical & Computer Engineering, Technion Israel Institute of Technology, Haifa 32000-03, Israel

[2]Swiss Light Source, Paul Scherrer Institute, CH-5232 Villigen, PSI, Switzerland

[a)]Author to whom correspondence should be addressed: brajagopal.technion@gmail.com

## ABSTRACT

Magnetic anisotropy in complex oxides often originates from the complex interplay of several factors, including crystal structure, spin-orbit coupling, and electronic interactions. Recent studies on Ru-substituted $La_{0.70}Sr_{0.30}MnO_3$ (Ru-LSMO) films demonstrate emerging magnetic and magneto-transport properties, where magnetic anisotropy plays a crucial role. However, the atomic origin and underlying mechanisms of the magnetic anisotropy of this material system remain elusive. This work sheds light on these aspects. Detailed element-specific X-ray magnetic dichroism analysis suggests that Ru single ion anisotropy governs the overall magnetic anisotropy. Furthermore, the magnetic property of Mn ions changes dramatically due to strong antiferromagnetic coupling between Ru and Mn ions. Our findings clarify the role of Ru single ion anisotropy behind magnetic anisotropy in Ru-LSMO, offering a promising avenue for designing advanced materials with tailored magnetic properties for next generation magnetic and spintronic technologies. As the Curie temperature of these materials is close to room temperature, such tunable magnetic anisotropy holds prospects for functional room-temperature magnetic devices.

## INTRODUCTION

Complex oxides provide a fertile playground for exploring and exploiting promising magnetic properties for next generation functional magnetic devices, owing to the intricate interplay of charge, spin, orbital, and lattice degrees of freedom.[1–9] In the last few years, Ru-substituted $La_{0.70}Sr_{0.30}MnO_3$ (Ru-LSMO) has garnered considerable attention because of its fascinating magnetic and magneto-transport properties, including a possible observation of skyrmions.[10,11] While $La_{0.70}Sr_{0.30}MnO_3$ (LSMO) films under slight compressive strain show in-plane magnetic anisotropy with a weak perpendicular magnetization component,[12–14] Ru substitution in the LSMO films strengthens perpendicular magnetization component under slight compressive strain.[10,11,15] A strong perpendicular magnetization is also observed in the related ruthenate $SrRuO_3$ (SRO), under similar strain.[4,5,16] These results suggest that 10%Ru-LSMO films preserve strong perpendicular magnetization of $SrRuO_3$ (SRO),[17] with the added benefit of the manganites' near-room temperature Curie temperature, suggesting a potentially similar underlying mechanism for magnetic anisotropy in both the systems.

Given the similarity with SRO, the overall magnetic anisotropy in Ru-LSMO is likely governed by either (i) single ion anisotropy of Ru ions,[16,18] or (ii) ligand (oxygen ion) driven anisotropy through orbital hybridization between O-2p and Ru-4d orbitals.[19,20] Unraveling the dominant mechanism of magnetic anisotropy in the Ru-LSMO films would not only advance the fundamental understanding of magnetic anisotropy, but further provide strategies for tailoring the magnetic properties of complex oxides for new functional devices, including magnetic memories, sensors, and spintronic devices.[8,9,21–24]. Moreover, as LSMO is a well-established double-exchange ferromagnet, it is essential to explore how Ru substitution alters the balance between exchange interactions and spin-orbit coupling, potentially leading to tunable magnetic properties.

The aim of this work is to elucidate the atomic origin of magnetic anisotropy in Ru-LSMO films and explain the atomic mechanism underlying the magnetic anisotropy. We report on an element-specific study of the magnetic properties along the principal (pseudo-cubic) crystallographic directions and discuss the atomic origin of the magnetic anisotropy and its underlying mechanisms. Our results suggest that Ru single ion anisotropy governs the overall magnetic anisotropy in the Ru-LSMO films. X-ray magnetic dichroism (XMCD) analysis suggests that the anisotropic orbital moments in the Ru ions lead to the single ion anisotropy through the strong spin-orbit coupling (SOC) on the Ru ions. The Ru spins then dictate the Mn spin through antiferromagnetic exchange interaction, resulting in the macroscopic magnetic anisotropy. This work thus clarifies the atomic origin and mechanism behind the magnetic anisotropy of Ru-LSMO films, highlighting how a small amount of transition metal ions with strong SOC can dramatically change the magnetic properties of complex oxide with intrinsically weak SOC.

## RESULTS AND DISCUSSION

The macroscopic magnetic anisotropy of crystalline thick (48nm) and thin (10.5nm) 10%Ru-LSMO was reported and explained in our previous work.[15] While both films exhibit tilted magnetic anisotropy (TMA) with strong perpendicular magnetization component, the in-plane magnetization is significantly anisotropic in the thick film. Structurally, both thick and thin Ru-LSMO films adopt monoclinic crystal structure with the octahedral tilt system $a^{+}a^{-}c^{-}$. In addition, the thick Ru-LSMO film features 1D periodic structural modulation, which further modifies the in-plane structure, resulting in significantly anisotropic in-plane magnetization.

The atomic mechanism of the macroscopic magnetic behavior of the LSMO films is generally determined by the B site ions of the perovskite structure ($ABO_3$), namely the Mn ions. In our Ru-LSMO films, 10% Mn is substituted by Ru ions. Therefore, to understand the atomic origin of the magnetic anisotropy in the 10%Ru-LSMO films, XMCD measurements were performed at Ru $M_{2,3}$-edges and Mn $L_{2,3}$-edges along the principal pseudo-cubic crystallographic directions (see details in the Experimental Section, Figure S1 and discussion therein). Each XMCD spectra was normalized by the corresponding area under the X-ray absorption spectra (XAS) for each direction.

The normalized XMCD and XAS spectra of Ru and Mn of Ru-LSMO are presented in Figure 1. We note three key features in these spectra: First, the Ru XMCD spectra indicate magnetic anisotropy in the Ru ions (Figure 1a), with the Ru $M_3$ peak of the out-of-plane (OOP) direction [001] being significantly larger than that of the in-plane direction [100] at high field (5T). Second, the Mn XMCD spectra (Figure 1c) are similar along these directions, indicating the magnetically isotropic behavior of the Mn ions at high field. Third, the positive (negative) Ru-$M_3$ ($M_2$) XMCD peak and negative (positive) Mn-$L_3$ ($L_2$) peak confirm that Ru ions are antiferromagnetically coupled with the Mn ions, in agreement with a previous observation.[25] Similar features are reproduced in the thin 10% Ru-LSMO film (Figure S2). While these qualitative observations reproduce our macroscopic picture of the films,[15] they are insufficient to understand the full picture behind the magnetic anisotropy of the 10% Ru-LSMO films; for this end we employ the systematic analysis of the sum rules and the magnetic field dependence of the Mn-XMCD.

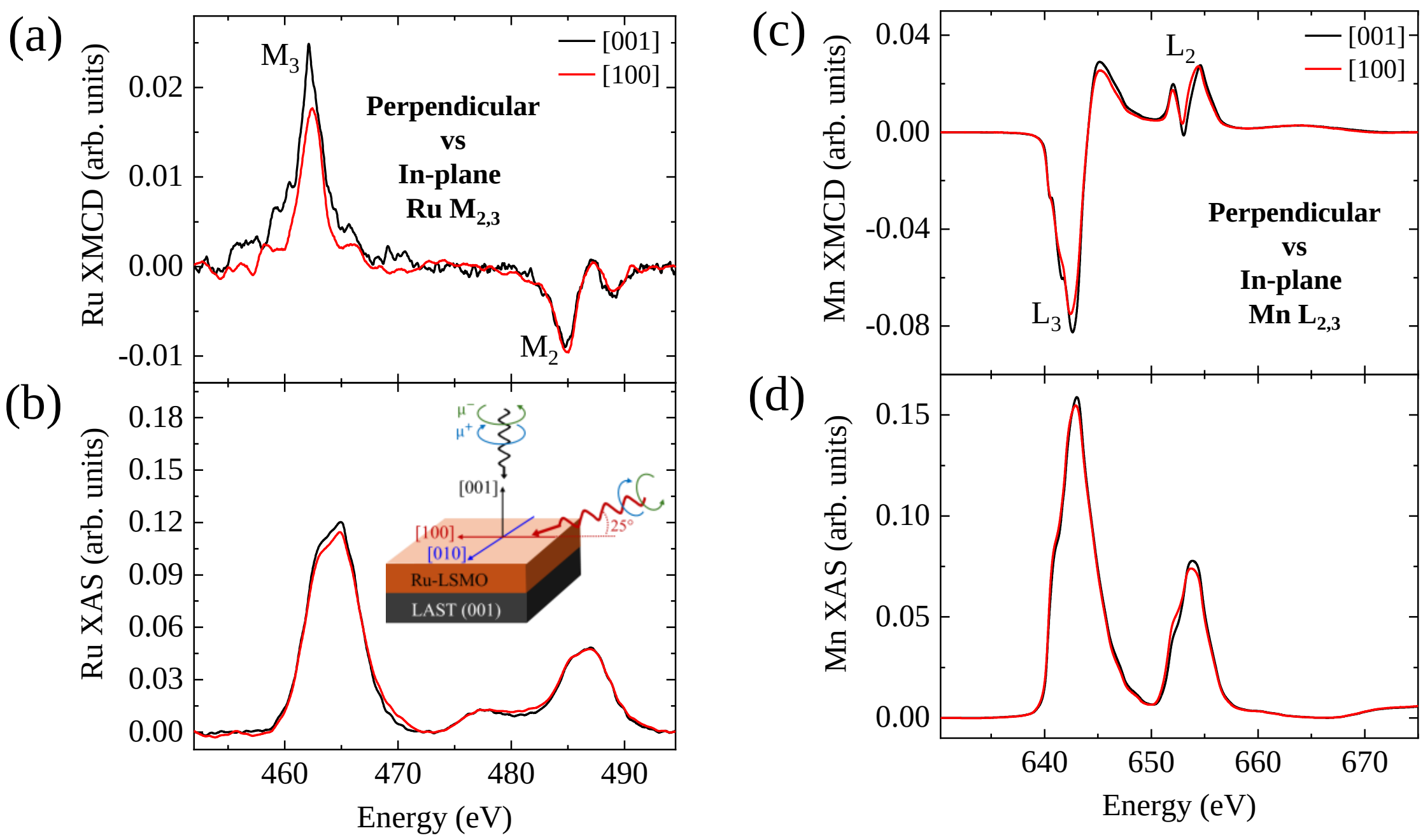

Figure 1. Normalized Ru $M_{2,3}$-edge (a) XMCD and (b) XAS of the thick (48nm) 10%Ru-LSMO film along the pseudo-cubic directions [001] and [100], normalization was done by the XAS area of Ru $M_{2,3}$ spectra for each direction. A schematic of the measurement axes is depicted in the inset. Normalized Mn $L_{2,3}$-edge (c) XMCD and (d) XAS of the thick 10%Ru-LSMO film, normalization was done by the XAS area of Mn $L_{2,3}$ spectra for each direction. All the Ru-$M_{2,3}$ and Mn-$L_{2,3}$ XMCD spectra were measured in ±5T, at 30K.

The sum rules were applied to these normalized XMCD spectra to calculate the orbital moment $(m_{orb})$ and effective spin moments $(m_{spin,eff})$ are as follows:[26–30]

$$m_{orb} = -(2/3)(A + B)(10 - n) \quad (1)$$

$$m_{spin,eff} = m_{spin} + 7m_T = -(A - 2B)(10 - n) \quad (2)$$

where A is area under the $M_3$ ($L_3$) peak of normalized Ru (Mn) XMCD spectra, B is area under the $M_2$ ($L_2$) peak of normalized Ru (Mn) XMCD spectra, n is the number of 4d (3d) electrons for Ru (Mn) ions, the term $m_{spin}$ indicates spin moment, and the term $m_T$ represents spin density distribution.[26]

To calculate the orbital and effective spin moments of the Ru ions, the value of n was taken as 4, considering predominantly the oxidation state of +4 in the Ru ions of Ru-LSMO films.[25] This implies that $Ru^{4+}$ ions replaces an equal proportions of the $Mn^{4+}$ ions to maintain the stoichiometry of the highly crystalline 10%Ru-LSMO films, resulting in 10% $Ru^{4+}$, 20% $Mn^{4+}$, and 70% $Mn^{3+}$ ions per formula unit, which leads to the average value of n as 3.8 for per Mn ion. Moreover, the spin moment of the Mn ions was corrected by a multiplying 1.5 due to admixture of the Mn $L_2$ and $L_3$ edges.[31,32]

The results of the sum rules are summarized in Table I. The key result is anisotropic behavior of the ratio of orbital moment to effective spin moment in Ru ions $\left(m_{orb}^{Ru}/m_{spin,eff}^{Ru}\right)$, confirming the presence of Ru single ion anisotropy. The key factor behind single ion anisotropy is SOC, which mediates the translation of anisotropic bonding environment into anisotropic orbital and spin moments.[26] The SOC in Ru ions is an order of magnitude larger than in Mn ions,[25] and the Ru 4d orbitals are more spatially extended than the Mn 3d orbitals.[11,33] As a result, the strong SOC of Ru ions amplifies the substrate-induced slight compressive strain into single ion anisotropy. In contrast, the weaker SOC of Mn ions is insufficient to induce this effect, resulting in negligible orbital moments, isotropic spin moment, and absence of single ion anisotropy in Mn ions. These results are consistent in the thin Ru-LSMO film as well (Figure S2 and discussion therein). This picture suggests that Ru single ion anisotropy is the driving force of magnetic anisotropy in the Ru-LSMO films. In addition, the orbital and spin moments of Ru ions are consistent with the previously reported values for Ru-LSMO, suggesting that the Ru ions adopt the +4 oxidation state.[25] As the XMCD spectra were acquired in surface sensitive total electron yield (TEY) mode (with a typical probing depth of ~5), the predominant +4 oxidation state of Ru ions and reasonably high values of spin moments of Mn ions suggest that there is no significant unwanted near surface region, reported in some complex oxide system.[34–37]

Table I. Orbital $(m_{orb})$ moments and effective spin $(m_{spin,eff})$ moments of Ru and Mn ions for [001] and [100] directions of the thick 10%Ru-LSMO film. The values were calculated using the sum rules [Eqs. (1), (2)] from high field (5T) data; all the units are in Bohr magneton.

| | | [001] | [100] |
|---|---|---|---|
| Ru ions | $m^{Ru}_{spin,eff}$ | -0.91 ± 0.03 | -0.69 ± 0.02 |
| | $m^{Ru}_{orb}$ | -0.25 ± 0.03 | -0.08 ± 0.01 |
| | $m^{Ru}_{total}$ | -1.16 ± 0.06 | -0.77 ± 0.03 |
| | $m^{Ru}_{orb}/m^{Ru}_{spin,eff}$ | 0.27 | 0.12 |
| Mn ions | $m^{Mn}_{spin,eff}$ | 3.04 ± 0.03 | 3.0 ± 0.03 |
| | $m^{Mn}_{orb}$ | -0.01 ± 0.01 | -0.01 ± 0.01 |
| | $m^{Mn}_{total}$ | 3.03 ± 0.04 | 2.99 ± 0.04 |

The results and discussion thus far focused on high fields (5T), where XAS was acquired (Figure 1). We now take a closer look at the Mn at lower fields. While the Mn ions exhibit isotropic magnetization (Fig. 1c, d, and Table I) at high field (5T) due to their fully aligned isotropic spin moments, a closer look at the magnetic field-dependent behavior (Figure 2) reveals the presence of TMA in Mn ions at lower fields, where perpendicular magnetization is stronger than in-plane magnetization. The XMCD hysteresis loops of the Mn ions show that TMA exists up to a threshold field of $\sim \pm 2.5$ T, with a strength of $1.6 \times 10^5$ J·m$^{-3}$ (consistent with the thin film, Figure S3). This strength is comparable with that of the overall TMA, which is $5.7 \times 10^5$ J·m$^{-3}$, calculated from macroscopic magnetization data acquired from these films.[15] On the other hand, in LSMO films (no Ru) under similar strain, the perpendicular magnetization component is negligible, resulting in nearly in-plane easy axis.[10,11,14] These results further underscore the important role of Ru single ion anisotropy in determining the overall magnetic anisotropy in Ru-LSMO films, by dictating the Mn spins through antiferromagnetic exchange interaction.

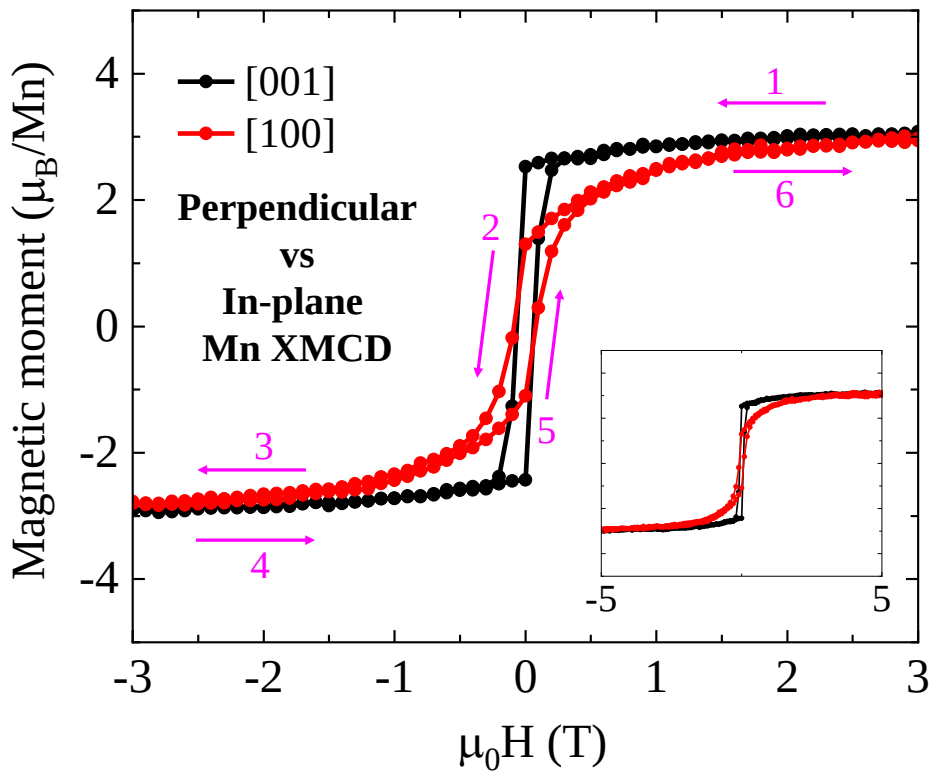

Figure 2. Magnetic field dependence of Mn XMCD along the [001] and [100] directions of the thick 10%Ru-LSMO film; the XMCD hysteresis loops are scaled using the Mn spin moments obtained from the sum rule analysis. The pink arrows indicate the sweeping direction, and the inset shows the full sweep range up to 5T. The measurements were performed at 30K.

The measured strength of TMA in Mn ions underscores the dominance of Ru-Mn antiferromagnetic coupling over Mn-Mn ferromagnetic coupling of pure LSMO, where in-plane magnetic anisotropy is observed under similar strain. This indicates that with only 10%Ru substitution, the Mn-Mn ferromagnetic interaction chain is significantly disrupted, causing magnetic easy axis of LSMO to shift from nearly in-plane direction to one that is closer to perpendicular direction. It suggests that Ru substitution not only alters local magnetic environment but also establishes a long range anisotropic magnetic order. Consequently, Ru single ion anisotropy rotates in-plane easy axis of LSMO towards the surface normal in Ru-LSMO, resulting in overall tilted magnetic anisotropy with strong perpendicular magnetization. These results further indicate that the strength of TMA in Mn ions depends on the strength of single ion anisotropy in Ru ions, as the Mn ions do not exhibit magneto-crystalline anisotropy. In other words, the strength of the single ion anisotropy in Ru ions is manifested in the TMA of the Mn ions. This implies that the strength of TMA in Mn ions should be the lower bound of the strength of Ru single ion anisotropy. Such a scenario is highly plausible, as the perpendicular magnetic anisotropy in Ru ions of SRO under similar strain is on the order of $10^6$ $J.m^{-3}$ (Refs. [4,5]) which scales to $10^5$ $J.m^{-3}$, when considering 10%Ru ions in Ru-LSMO films.

Achieving a similar strength of TMA with strong perpendicular magnetization component in LSMO films (no Ru) requires high compressive strain ($\sim -2\%$).[38,39] This underscores the efficacy of Ru single ion anisotropy in the Ru-LSMO films, paving the way for designing materials with tailored magnetic properties under low strain, which is desirable for practical device applications. Tilted magnetic anisotropy with strong perpendicular magnetization offers significant advantages over in-plane magnetic anisotropy for spintronic applications.[9,22,40,41] This finding thus highlights a unique approach to introducing TMA with strong perpendicular magnetization in materials with high Curie temperature, such as manganites, achievable through a small substitution of an element with high SOC.

Given the relevance of SRO and other complex oxides to the observed anisotropy, we now discuss two additional potential contributors – oxygen ligand and oxygen vacancy – and rule out their significant contributions. If the ligand-oxygen-driven magnetic anisotropy, recently reported in SRO,[20] were dominant in Ru-LSMO, then the ratio of orbital moment to effective spin moment in Ru ions would be isotropic. However, the significantly anisotropic ratio in Ru ions of Ru-LSMO (Table I) suggests that Ru single ion anisotropy is the dominant factor, over the possibility of oxygen ligand driven anisotropy. The possibility of any significant contribution from oxygen vacancy driven anisotropy[42,43] can be ruled out based on the three key factors: the overall high structural quality reported in our previous study of these films,[15] the predominant +4 oxidation state of Ru ions, and reasonably high values of spin moments of Mn ions.[25]

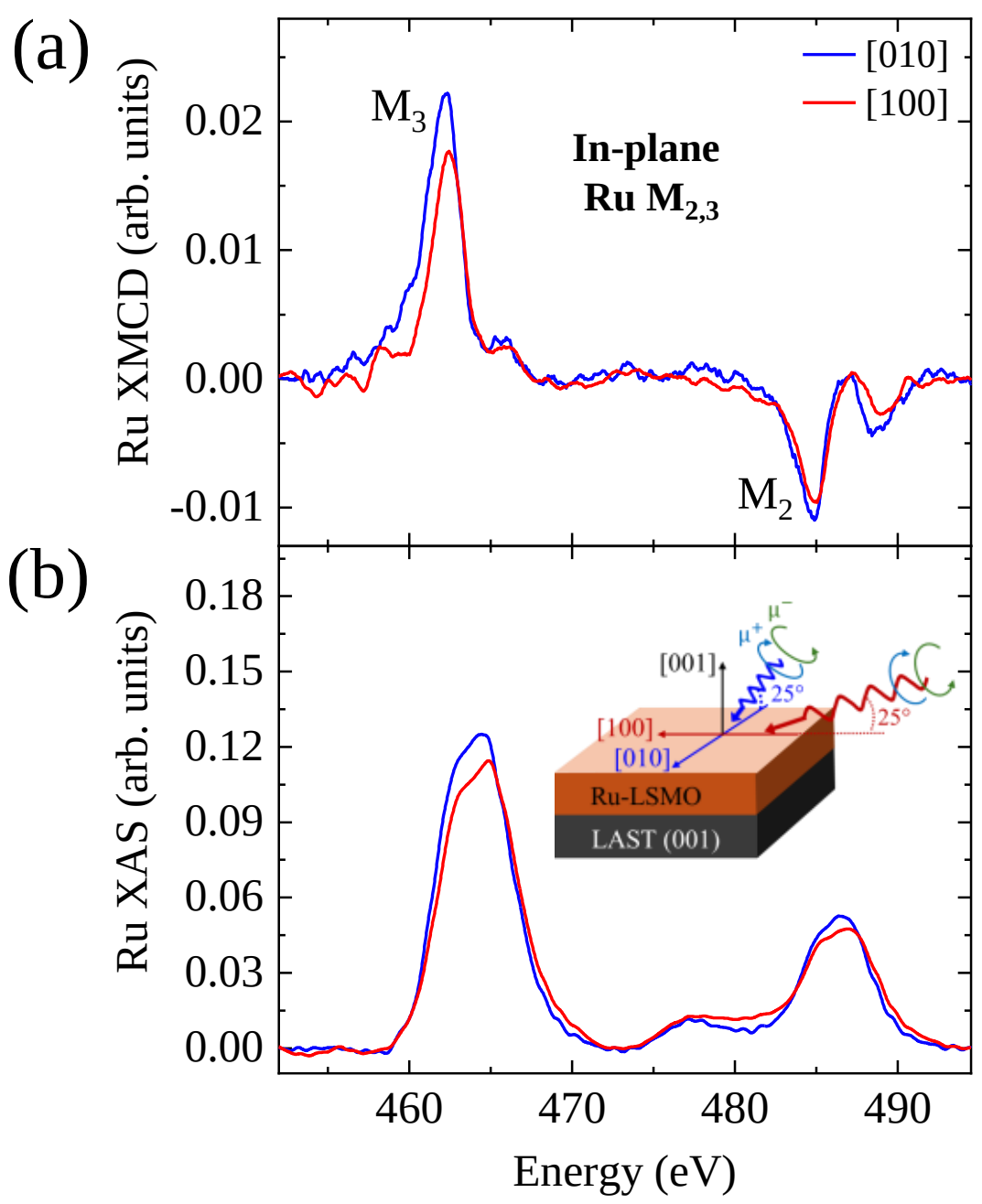

Figure 3. Normalized Ru $M_{2,3}$-edge (a) XMCD and (b) XAS of the thick (48nm) 10%Ru-LSMO film along the in-plane pseudo-cubic directions [100] and [010] at 30K; normalization was done by the corresponding total absorption for each direction. The schematic of the measurement set up is depicted in the inset. All the Ru-$M_{2,3}$ and Mn-$L_{2,3}$ XMCD spectra were measured in ±5T, at 30K.

We now briefly discuss the atomic origin of anisotropic in-plane magnetization in the thick Ru-LSMO film. As demonstrated in our previous work, 1D periodic structural modulation emerges above a critical thickness due to shear strain relaxation, resulting in significantly anisotropic in-plane magnetization.[15] This anisotropic in-plane bonding environment should introduce anisotropic orbital moments, anisotropic spin moments, and anisotropic ratio of orbital moments to effective spin moments at the atomic level.[26] This effect is precisely observed in the in-pane XMCD spectra of Ru ions (Figure 3 and Table II). These results suggest that Ru single ion anisotropy (Figure 4) is not only responsible for the emergence of TMA but also plays the key role in driving anisotropic in-plane magnetization. This further reinforces the key role of Ru single ion anisotropy behind the overall magnetic anisotropy.

Table II. Orbital $(m_{orb})$ moments and effective spin $(m_{spin,eff})$ moments of Ru and Mn ions for [010] and [100] directions of the thick 10%Ru-LSMO film were calculated using the sum rules; all the units are in Bohr magneton.

| | | [010] | [100] |
|---|---|---|---|
| Ru ions | $m^{Ru}_{spin,eff}$ | -0.85 ± 0.02 | -0.69 ± 0.02 |
| | $m^{Ru}_{orb}$ | -0.17 ± 0.02 | -0.08 ± 0.01 |
| | $m^{Ru}_{total}$ | -1.02 ± 0.04 | -0.77 ± 0.03 |
| | $m^{Ru}_{orb}/m^{Ru}_{spin,eff}$ | 0.2 | 0.12 |

We further note that the single ion anisotropy (Figure 4) in Ru ions has two components: anisotropy in orbital moment and anisotropy in effective spin moment. While both these anisotropy stem from anisotropic bonding environment, the later typically contributes as a second order effect to the overall magnetic anisotropy.[26,44] The precise contribution of each component requires further investigation. Nevertheless, here the anisotropic bonding environment is driven by substrate-induced strain. While the compressive strain gives rise to anisotropy between in-plane and out-of-plane directions by compressing unit cells in the plane and elongating it in the out-of-plane direction, the anisotropic shear strain relaxation gives rise to in-plane anisotropy by forming 1D periodic structural modulation above a critical thickness.[15]

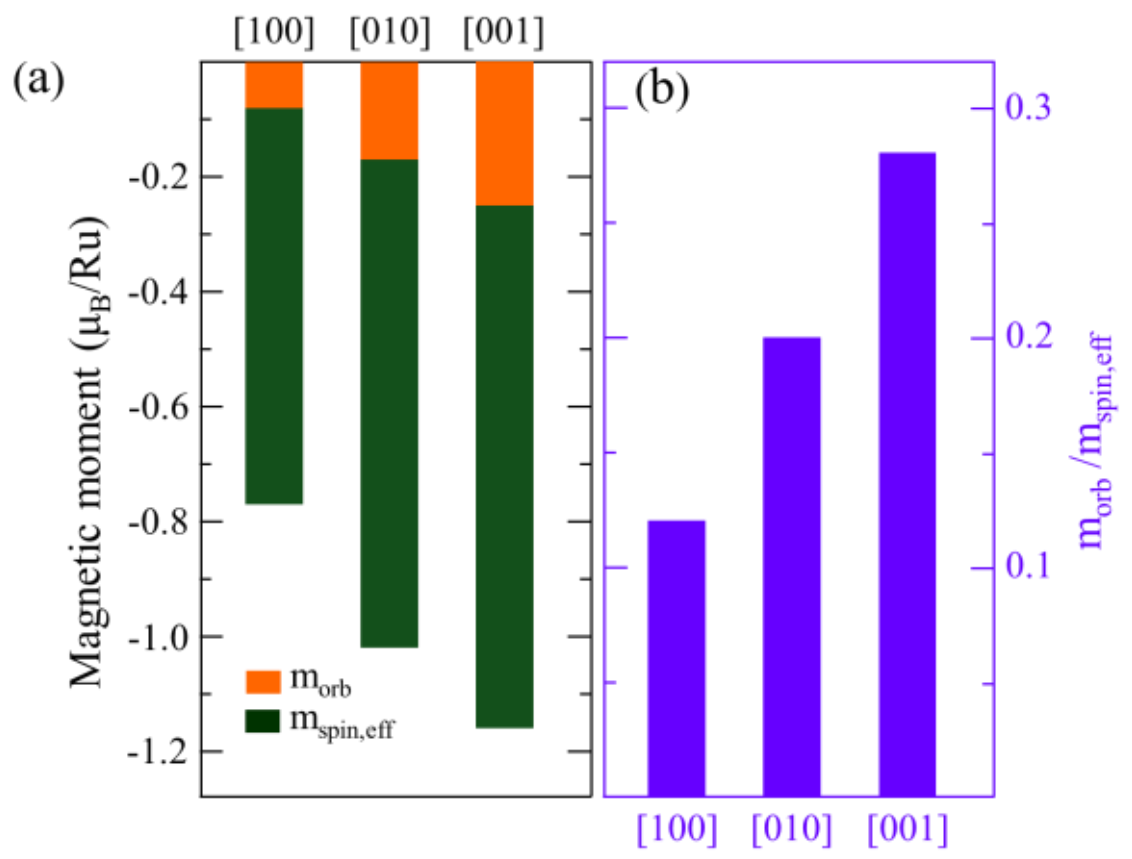

Figure 4. Ru single ion anisotropy of the thick 10%Ru-LSMO film, showing (a) orbital moments, effective spin moments, and (b) the ratio of orbital moment to effective spin moment along the three main pseudo-cubic directions.

Overall, the results suggest that substrate induced strain introduces single ion anisotropy in Ru ions through strong SOC. This Ru single ion anisotropy subsequently dictates the orientation of Mn spins through antiferromagnetic exchange interaction. Moreover, TMA with strong perpendicular magnetization in Ru-LSMO films suggests that Ru-Mn antiferromagnetic exchange interaction dominates over Mn-Mn ferromagnetic exchange interaction, as LSMO films exhibit in-plane anisotropy under similar strain. In this way, Ru single ion anisotropy establishes an anisotropic long range magnetic order, ultimately giving rise to overall magnetic anisotropy in Ru-LSMO films.

## SUMMARY AND CONCLUSION

The atomic origins of magnetic anisotropy and the underlying atomic mechanism in Ru-LSMO films were elucidated using element-specific XMCD. Our systematic analysis reveals that Ru single ion anisotropy, induced by substrate induced strain through strong SOC, governs the overall magnetic anisotropy by dictating Mn spins through Ru-Mn antiferromagnetic coupling. These findings highlight how a small fraction of elements with high SOC can dramatically alter the magnetic properties of 3d transition metal-based complex oxides, by leveraging high SOC instead of relying solely on strain, providing a better alternative to strain-based approach. This understanding offers a promising pathway for designing advanced materials with tailored magnetic properties for next generation magnetic and spintronic technologies.

EXPERIMENTAL

Thin film growth: Ru-LSMO films were epitaxially grown on LSAT substrates using pulsed laser deposition (PLD) at a substrate temperature of 650 °C, KrF laser fluence of 2.4 J $cm^{-2}$, and a repetition rate of 3-5 Hz at ~0.13 mbar of $O_2$ (see a previous work[15] for the full details).

X-ray magnetic dichroism: X-ray absorption spectroscopies (XAS) of a thick (48nm) and a thin (10.5nm) 10% Ru-LSMO films were acquired with right- ($\mu^+$) and left-handed ($\mu^-$) circularly polarized light at the X-Treme beamline of Swiss Light Source. The spectra were taken along the principal crystallographic (pseudocubic) directions. For the [001] direction, the x-rays incident perpendicular to the film surface; for the in-plane measurements, the x-rays incident at a grazing angle of 25° with the pseudocubic crystallographic direction [100] ([010]) in the xz (yz) plane. The XAS is defined as the arithmetic average of intensities $[I(\mu^+) + I(\mu^-)]/2$ of the measured spectra with right ($\mu^+$) and left ($\mu^-$) circularly polarized light. The XMCD is the difference $(I(\mu^+) - I(\mu^-))$ normalized by the corresponding total absorption, meaning $\mathrm{XMCD} = (I(\mu^+) - I(\mu^-))\ /(\mathrm{XAS\ area})$. Along each crystallographic direction, the final XMCD results were obtained from the average of the two normalized XMCD spectra, with the externally applied magnetic field of 5T parallel and anti-parallel to the direction of propagation of x-rays. All the XMCD spectra were acquired in total electron yield (TEY) mode, and the XMCD hysteresis loops were acquired in X-ray excited optical luminescence (XEOL) mode. While TEY is a surface sensitive technique, XEOL is a transmission-based bulk sensitive technique probing entire film.[36,37,45] All the measurements were performed at 30K.

ACKNOWLEDGEMENT

This work was funded by the German Israeli Foundation (GIF Grant no. I-1510-303.10/2019) and the Israeli Science Foundation (ISF Grant No. 1351/21). The authors thank Dr. Ionela Lindfors-Vrejoiu (University of Cologne) for expertly growing the films used here. The authors further thank Prof. Paul H.M. van Loosdrecht (University of Cologne) and Prof. Robert Green (University of Saskatchewan) for fruitful discussions.

# SUPPLEMENTARY MATERIAL

## Atomic Origins of Magnetic Anisotropy in Ru-substituted Manganite Films

Brajagopal Das,[1] Cinthia Piamonteze,[2] Lior Kornblum[1]

[1]Andrew & Erna Viterbi Department of Electrical & Computer Engineering, Technion Israel Institute of Technology, Haifa 3200003, Israel

[2]Swiss Light Source, Paul Scherrer Institute, CH-5232 Villigen, PSI, Switzerland

**Background correction of the Ru-$M_{2,3}$ and Mn-$L_{2,3}$ spectra of the thick (48nm) 10%Ru-LSMO film**

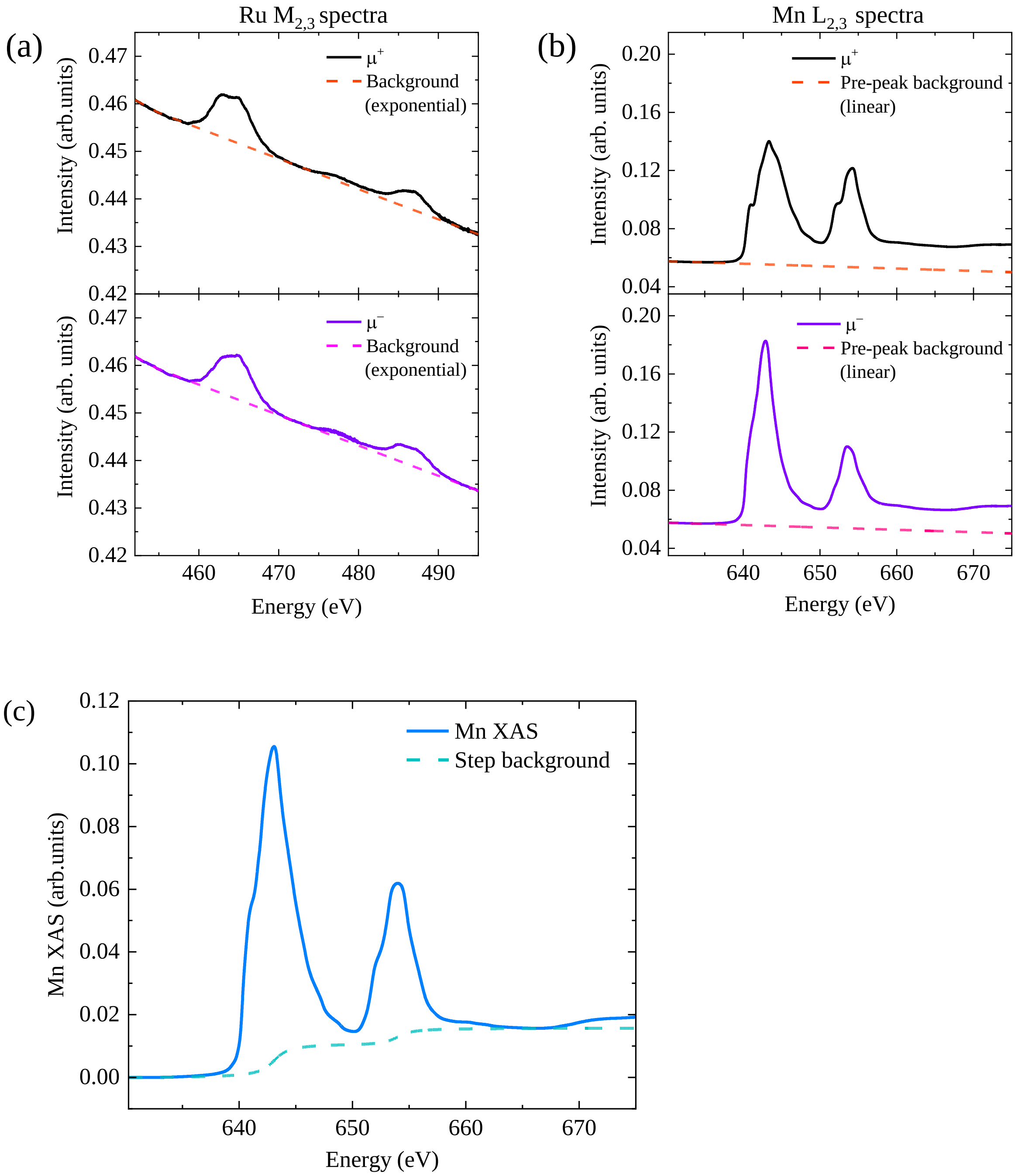

Figure S1. (a) Ru $M_{2,3}$ spectra (solid curves, raw) with exponential (piecewise) background (dashed curves, fitted), and (b) Mn $L_{2,3}$ spectra (solid curves, raw) with pre-peak linear background (dashed curves, fitted) for right ($\mu^+$) and left ($\mu^-$) circularly polarized x-rays. (c) Mn XAS plot with edge jumps at the $L_{2,3}$ edges, where XAS is defined as the average of intensities I($\mu^+$) and I($\mu^-$), after pre-peak linear background correction. All the measurements were performed at 30K.

The shape of the raw Ru $M_{2,3}$ spectra and the fitted background for Ru $M_{2,3}$ spectra agree well with the earlier XMCD study of Ru ions in Ru-LSMO films.[1] The Ru $M_{2,3}$ spectra with intensities $I(\mu^+)$ and $I(\mu^-)$ were obtained after the exponential background subtraction. These spectra were subsequently used for XMCD $(I(\mu^+) - I(\mu^-))$ and XAS $([I(\mu^+) + I(\mu^-)]/2)$ plots. For Mn $L_{2,3}$ spectra, pre-peak linear background has been fitted and then the straight line has been extrapolated well beyond the $L_2$ edge. After the linear background subtraction, the Mn $L_{2,3}$ spectra of $I(\mu^+)$ and $I(\mu^-)$ were used for XMCD and XAS; to apply the orbital and spin sum rules, the edge jump background with the jumps at the Mn $L_3$ and $L_2$ edges was subtracted from Mn XAS, as described elsewhere.[2] The procedure of background subtraction for Ru and Mn ions explained here is for [001] direction of the thick Ru-LSMO film; however, the same procedures were applied to all the crystallographic directions of both the thick and thin 10%Ru-LSMO films.

**Element specific magnetic anisotropy in the thin (10.5nm) 10%Ru-LSMO film**

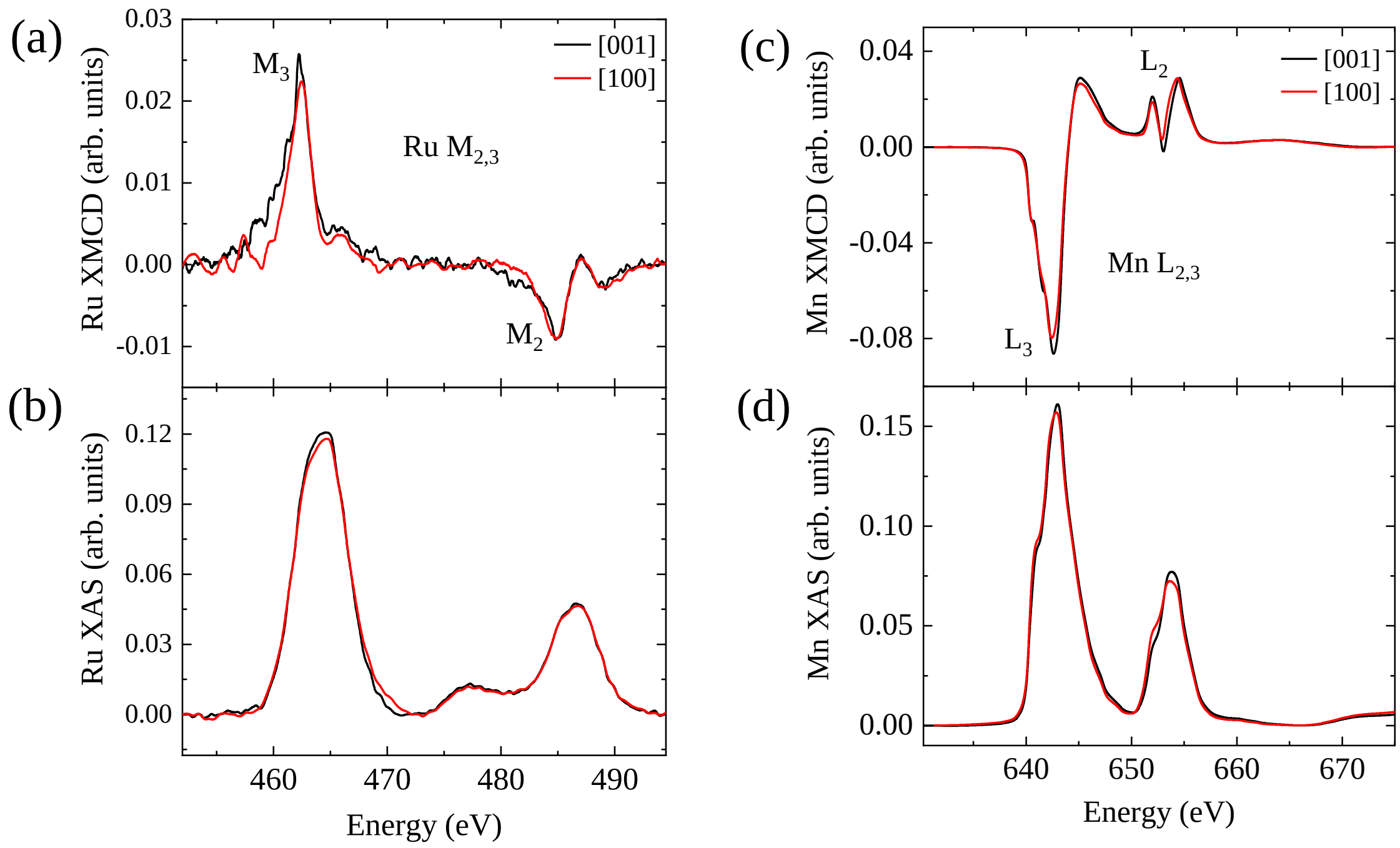

Figure S2. Normalized Ru $M_{2,3}$-edge (a) XMCD and (b) XAS of the thin (10.5 nm) 10%Ru-LSMO film along the pseudo-cubic directions [001] and [100], normalization was done by the XAS area of Ru $M_{2,3}$ spectra for each direction. Normalized Mn $L_{2,3}$-edge (c) XMCD and (d) XAS of the thin 10%Ru-LSMO film, normalization was done by the XAS area of Mn $L_{2,3}$

spectra for each direction. All the Ru-$M_{2,3}$ and Mn-$L_{2,3}$ XMCD spectra were measured in ±5T, at 30K.

The normalized XMCD and XAS spectra of Ru and Mn of the thin Ru-LSMO film are presented in Figure S2. We note three key features in these spectra: First, the Ru XMCD spectra indicate magnetic anisotropy in the Ru ions (Figure S2a), with the Ru $M_3$ peak of the out-of-plane (OOP) direction [001] being significantly larger than that of the in-plane direction [100] at high field (5T). Second, the Mn XMCD spectra (Figure S2c) are similar along these directions, indicating the magnetically isotropic behavior of the Mn ions at high field. Third, the positive (negative) Ru-$M_3$ ($M_2$) XMCD peak and negative (positive) Mn-$L_3$ ($L_2$) peak confirm that Ru ions are antiferromagnetically coupled with the Mn ions, in agreement with a previous observation.[1] While these qualitative observations reproduce our macroscopic picture of the films,[3] they are insufficient to understand the full picture behind the magnetic anisotropy of the 10% Ru-LSMO films; for this end we employ the systematic analysis of the sum rules and the magnetic field dependence of the Mn-XMCD. In addition, we note that the small Ru-XAS peak (Figure S2b) around 477eV originates due to the transition from 3p to 5s orbitals of the Ru ions, with no magnetic contribution as observed in the XMCD spectra (Figure S2a).[4,5]

The results of the sum rules are summarized in Table SI. The key result is anisotropic behavior of the ratio of orbital moment to effective spin moment in Ru ions $\left(m_{orb}^{Ru}/m_{spin,eff}^{Ru}\right)$, confirming the presence of Ru single ion anisotropy. The key factor behind single ion anisotropy is SOC, which mediates the translation of anisotropic bonding environment into anisotropic orbital and spin moments.[6] The SOC in Ru ions is an order of magnitude larger than in Mn ions,[1] and the Ru 4d orbitals are more spatially extended than the Mn 3d orbitals.[7,8] As a result, the strong SOC of Ru ions amplifies the substrate-induced slight compressive strain into single ion anisotropy. In contrast, the weaker SOC of Mn ions is insufficient to induce this effect, resulting in negligible orbital moments, isotropic spin moment, and absence of single ion anisotropy in Mn ions. This picture suggests that Ru single ion anisotropy is the driving force of magnetic anisotropy in the Ru-LSMO films. In addition, the orbital and spin moments of Ru ions are consistent with the previously reported values for Ru-LSMO, suggesting that the Ru ions adopt +4 oxidation state.[1] As the XMCD spectra were acquired in surface sensitive TEY mode (with a typical probing depth 5 – 10 nm), the predominant +4 oxidation state of Ru ions and reasonably high values of spin moments of Mn ions suggest that there is no significant unwanted near surface region, reported in some complex oxide system.[9–12]

Table SI. Orbital ($m_{orb}$) moments and effective spin ($m_{spin,eff}$) moments of Ru and Mn ions for [001] and [100] directions of the thin 10%Ru-LSMO film were calculated using the sum rules; all the units are in Bohr magneton.

| | | [001] | [100] |
|---|---|---|---|
| Ru ions | $m^{Ru}_{spin,eff}$ | -0.94 ± 0.02 | -0.75 ± 0.02 |
| | $m^{Ru}_{orb}$ | -0.25 ± 0.03 | -0.15 ± 0.02 |
| | $m^{Ru}_{total}$ | -1.19 ± 0.05 | -0.90 ± 0.04 |
| | $m^{Ru}_{orb}/m^{Ru}_{spin,eff}$ | 0.27 | 0.2 |
| Mn ions | $m^{Mn}_{spin,eff}$ | 3.2 ± 0.04 | 3.16 ± 0.03 |
| | $m^{Mn}_{orb}$ | -0.01 ± 0.01 | -0.01 ± 0.01 |
| | $m^{Mn}_{total}$ | 3.19 ± 0.05 | 3.15 ± 0.04 |

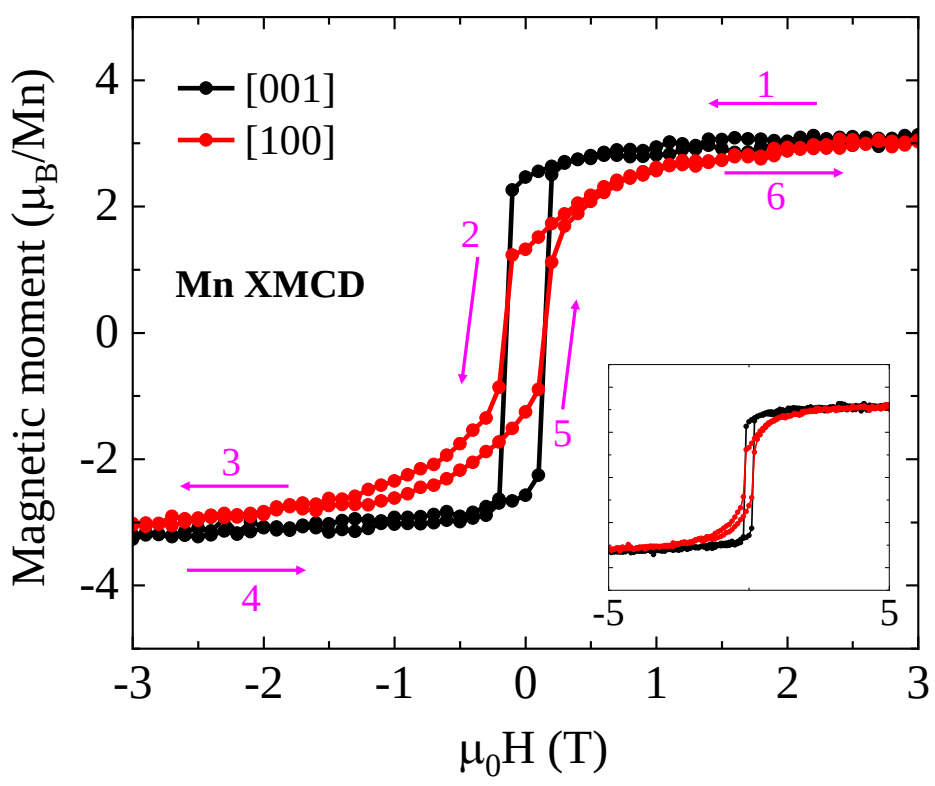

Figure S3. Magnetic field dependence of Mn XMCD along the [001] and [100] directions of the thin 10%Ru-LSMO film; the XMCD hysteresis loops are scaled using the Mn spin moments obtained from the sum rule analysis. The pink arrows indicate the sweeping direction, and the inset shows the full sweep range up to 5T. The measurements were performed at 30K.

While the Mn ions exhibit isotropic magnetization (Fig. S2c and Table SI) at high field (5T) due to their fully aligned isotropic spin moments, a closer look at the magnetic field-dependent behavior (Figure S3) reveals the presence of TMA in Mn ions at lower fields, where perpendicular magnetization is stronger than in-plane magnetization. The XMCD hysteresis loops of the Mn ions show that TMA exists up to a threshold field of $\sim \pm 2.5$ T, with a strength of $1.1 \times 10^5$ J·m$^{-3}$ This strength is comparable with that of the overall TMA, which is $4 \times 10^5$ J·m$^{-3}$, calculated from macroscopic magnetization data acquired from the thin Ru-LSMO film.[3] On the other hand, in LSMO films (no Ru) under similar strain, the perpendicular magnetization component is negligible, resulting in nearly in-plane easy axis.[7,13,14] These results further underscore the important role of Ru single ion anisotropy in determining the overall magnetic anisotropy in Ru-LSMO films, by dictating the Mn spins through antiferromagnetic exchange interaction.